\documentclass[twocolumn,showpacs,preprintnumbers,amsmath,amssymb]{revtex4}

\usepackage{graphicx}
\usepackage{dcolumn}
\usepackage{bm}
\usepackage{color}

\begin{document}

\preprint{APS/123-QED}

\title{Novel critical behavior of magnetization in URhSi:\\Similarities to uranium ferromagnetic superconductors UGe$_2$ and URhGe\footnote{Phys. Rev. B {\bf 99}, 094417 (2019).}}

\author{Naoyuki Tateiwa$^{1}$}
\email{tateiwa.naoyuki@jaea.go.jp} 
\author{Yoshinori Haga$^{1}$}
\author{Etsuji Yamamoto$^{1}$}%

\affiliation{
$^{1}$Advanced Science Research Center, Japan Atomic Energy Agency, Tokai, Naka, Ibaraki 319-1195, Japan\\
}
\date{\today}

\begin{abstract}
We study the critical behavior of dc magnetization in the uranium ferromagnet URhSi around the paramagnetic to ferromagnetic phase transition at $T_{\rm C}{\,}{\sim}{\,}$10 K with a modified Arrott plot, a Kouvel-Fisher plot, the critical isotherm analysis and the scaling analysis. URhSi is isostructural to uranium ferromagnetic superconductors URhGe and UCoGe. The critical exponent $\beta$ for the temperature dependence of the spontaneous magnetization below $T_{\rm C}$, $\gamma$ for the magnetic susceptibility, and $\delta$ for the magnetic isotherm at $T_{\rm C}$ in URhSi have been determined as $\beta$ = 0.300 $\pm$ 0.002, $\gamma$ = 1.00 $\pm$ 0.02, and $\delta$ = 4.38 $\pm$ 0.04 by the scaling analysis and the critical isotherm analysis. These critical exponents fulfill the Widom scaling law ${\delta}{\,}={\,}1+{\,}{\gamma}/{\beta}$. Magnetization has strong uniaxial magnetic anisotropy in the ferromagnetic state of URhSi. However, the universality class of the ferromagnetic transition does not belong to the 3D Ising system with short-range exchange interactions between magnetic moments ($\beta$ = 0.325, $\gamma$ = 1.241, and $\delta$ = 4.82). The obtained exponents in URhSi are similar to those in the uranium ferromagnetic superconductors UGe$_2$ and URhGe, and uranium ferromagnets UIr and U(Co$_{0.98}$Os$_{0.02}$)Al. We have previously reported the unconventional critical behavior of magnetization in the uranium ferromagnetic superconductors [N. Tateiwa {\it et al.} Phys. Rev. B {\bf 89}, 064420 (2014)]. The universality class of the ferromagnetic transition in URhSi may belong to the same one in the uranium ferromagnetic superconductors and the uranium ferromagnets. The unconventional critical behavior of the magnetization in the uranium compounds cannot be understood with previous theoretical interpretations of critical phenomena. The absence of the superconductivity in URhSi is discussed from several viewpoints. The improvement of the sample quality in URhSi could provide a good opportunity to gain a deeper understanding of the ferromagnetic superconductivity in the uranium ferromagnets. 
\end{abstract}

\maketitle

\section{Introduction}
\subsection{General introduction}
Many experimental and theoretical studies have been done for intriguing physical properties in uranium compounds with $5f$ electrons such as mysterious ``hidden order" in URu$_2$Si$_2$, unconventional superconductivity in UPt$_3$ or UBe$_{13}$, and the coexistence of the superconductivity and antiferromagnetism in UPd$_2$Al$_3$ or UNi$_2$Al$_3$\cite{mydosh,stewart,pfleiderer0}. The most unique feature of uranium $5f$ systems is the coexistence of the superconductivity and ferromagnetism both carried by the same $5f$ electrons in UGe$_2$, URhGe, and UCoGe\cite{saxena,huxley1,aoki0,huy,pfleiderer0,huxley2}. Novel physical phenomena associated with a quantum phase transition between ferromagnetic and paramagnetic states have been the subjects of extensive researches from both experimental and theoretical sides\cite{brando}. 

  It is important to understand detailed ferromagnetic properties in the uranium ferromagnetic superconductors UGe$_2$, URhGe, and UCoGe for a better understanding of the superconductivity. This is because ferromagnetic interactions between the $5f$ electrons may play an important role for the appearance of the superconductivity in the ferromagnetic state as theoretically shown\cite{fay,roussev,kirkpatrick,wang}. The ferromagnetic states in UGe$_2$, URhGe, and UCoGe are magnetically uniaxial\cite{sakon,hardy,huy2}. The uranium ferromagnetic superconductors have been regarded as a three-dimensional (3D) Ising system. We focus on a classical critical behavior of the magnetization around a ferromagnetic transition temperature from which the type of the magnetic phase transition and the nature of magnetic interactions can be studied\cite{privman}. We have previously reported that the universality class of the critical phenomena in UGe$_2$ and URhGe does not belong to any known universality classes of critical phenomena\cite{tateiwa1}. The ferromagnetism of the uranium ferromagnetic superconductors may not be described only with the 3D Ising model.

In this paper, we report the novel critical behavior of the magnetization in URhSi. The compound crystalizes in the same orthorhombic TiNiSi-type crystal structure (space group $Pnma$) to those of the uranium ferromagnetic superconductors URhGe and UCoGe\cite{sechovsky}. URhSi shows a ferromagnetic transition at the Curie temperature $T_{\rm C}{\,}{\sim}{\,}$10 K. The superconductivity has not been observed down to 40 mK\cite{prokes2}. The ferromagnetic state in URhSi has uniaxial magnetic anisotropy with the magnetic easy axis parallel to the $c$ axis in the orthorhombic crystal structure\cite{prokes2}, which is similar to the uranium ferromagnetic superconductors URhGe\cite{hardy} and UCoGe\cite{huy2}. We find that the universality class of the critical phenomenon in URhSi does not belong to the 3D Ising model. The values of the critical exponents in URhSi are similar to those in URhGe and UGe$_2$. The universality class of the ferromagnetic transition in URhSi may belong to the same one in URhGe and UGe$_2$. We discuss the static and dynamical magnetic properties of URhSi in comparison with those of URhGe, UCoGe, and UGe$_2$. Possible reasons for the absence of the superconductivity in URhSi are discussed.

 \subsection{Physical properties in URhSi and comparison with URhGe, UCoGe, and UGe$_2$}
We summarize the crystal structure and basic physical properties of URhSi in this subsection. Figure 1 shows the orthorhombic TiNiSi-type crystal structure and Table I shows the structural parameters of URhSi. Here, $B_{\rm eq}$ is the equivalent isotropic atomic displacement parameter. Lattice parameters at room temperature are determined as $a$ = 0.69970(4) nm, $b$ = 0.42109(2) nm, and $c$ = 0.74458(4) nm by single-crystal x-ray diffraction techniques using an imaging plate area detector (Rigaku) with Mo K${\alpha}$ radiation. The distances between the uranium atoms are $d_1$ = 0.3638 nm and $d_2$ = 0.3408 nm along the $a$ and $b$ axes, respectively. The structure can be regarded as coupled chains of the nearest-neighbor uranium atoms (``zigzag chain") running along the crystallographic $b$ axis. Meanwhile, the distance $d_1$ is shorter than $d_2$ in URhGe and UCoGe\cite{pfleiderer0}. The crystal structure of both compounds can be viewed as the coupled zigzag chains along the $a$ axis. 

Table II tabulates the values of $T_{\rm C}$, $p_{\rm eff}$, $p_{\rm s}$, $T_0$, $T_{\rm A}$, and ${T_{\rm C}}/{T_{\rm 0}}$ for URhSi, URhGe, UCoGe, and UGe$_2$. Here, $p_{\rm eff}$ and $p_{\rm s}$ are the effective and the spontaneous magnetic moments, respectively. The definitions of $T_0$ and $T_{\rm A}$ will be explained later. The values of the parameters for URhSi are determined in this study. UGe$_2$ orders ferromagnetically at the relatively high Curie temperature $T_{\rm C}$ of 52.6 K with the large spontaneous magnetic moment $p_{\rm s}$ = 1.41 ${\mu}_{\rm B}$/U\cite{tateiwa1}. Meanwhile, UCoGe shows the ferromagnetic transition at $T_{\rm C}$ = 2.4 K with the small spontaneous magnetic moment $p_{\rm s}$ = 0.0039 ${\mu}_{\rm B}$/U\cite{huy}. The values of $T_{\rm C}$ and $p_{\rm s}$ in URhSi are similar to those in URhGe\cite{tateiwa1}. Neutron diffraction studies on URhSi have shown a collinear ferromagnetic structure with an uranium magnetic moment of 0.50 - 0.55 ${\mu}_{\rm B}$/U oriented along the $c$ axis\cite{tran,prokes1}. The ferromagnetic structure is the same as those in URhGe and UCoGe\cite{aoki0, prokes4}. The linear specific heat coefficient $\gamma$ in URhSi was determined as $\gamma$ = 164.2 mJ/(mol$\cdot$K$^2$) in the ferromagnetic ordered state\cite{prokes2}. This value is almost the same as that [$\sim$ 160 mJ/(mol$\cdot$K$^2$)] in URhGe estimated from the $C/T$ value just above the superconducting transition temperature $T_{\rm sc}$\cite{sakarya2}. The $\gamma$ value in UCoGe [= 57 mJ/(mol$\cdot$K$^2$)] is about one-third of those in URhSi and UCoGe\cite{huy}. There are several similarities in the basic physical properties between URhSi and UCoGe. 

   \begin{figure}[t]
\includegraphics[width=8cm]{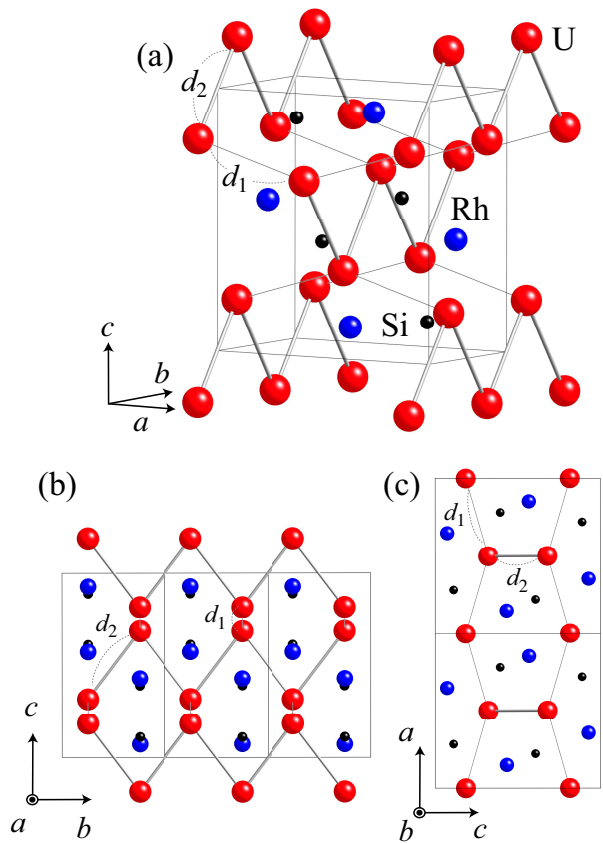}
\caption{\label{fig:epsart}(a) Representation of the orthorhombic TiNiSi-type crystal structure of URhSi. Projections of atoms on (b) the $bc$ and (c) the $ac$ planes. }
\end{figure}

 \begin{table}[]
\caption{\label{tab:table1}%
Crystallographic parameters for URhSi at room temperature in the orthorhombic setting (space group $Pnma$) with lattice parameters $a$ = 0.69970(4) nm, $b$ =  0.42109(2) nm, and $c$ =  0.74458(4) nm. The conventional unweighted and weighted agreement factors of $R_1$ and $wR_2$ are 3.47 and 9.01$\%$, respectively.}
\begin{ruledtabular}
\begin{tabular}{cccccc}
\textrm{Atom}&
\textrm{Site}&
\textrm{$x$}&
\textrm{$y$ }&
\textrm{$z$}&
\textrm{$B_{\rm eq}$ (nm$^2$)}\\
\colrule
U&4(c) & 0.00257(6)  &1/4 & 0.18536(7)   &$5.1(2){\times}10^{-3}$\\ 
Rh &4(c)   & 0.15004(17) & 1/4&0.57177(14)  &$7.7(3){\times}10^{-3}$\\ 
Si&4(c)  &0.7870(7) &1/4 &0.6056(5) &$5.7(6){\times}10^{-3}$ \\ 
\end{tabular}
\end{ruledtabular}
 \end{table}
 
   \begin{figure}[]
\includegraphics[width=8 cm]{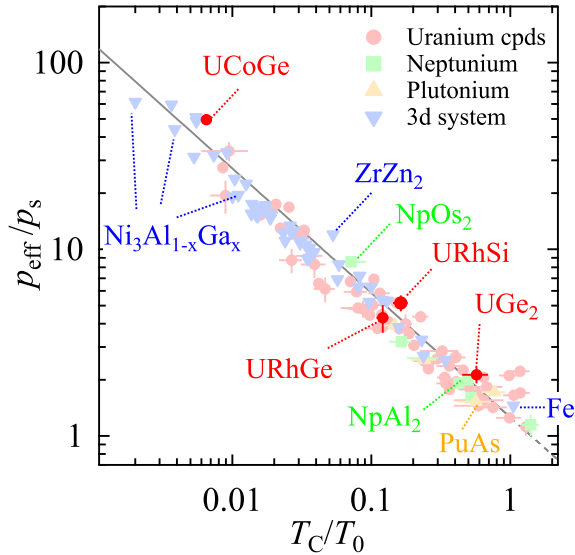}
\caption{\label{fig:epsart}Generalized Rhodes-Wohlfarth plot for uranium, neptunium and plutonium ferromagnets, and the $3d$ metals and their intermetallic ferromagnetic compounds shown as closed circles, squares, triangles and anti-triangles, respectively\cite{tateiwa2}. The data for UCoGe and the $3d$ systems are cited from the literature\cite{nksato1,takahashi1,takahashi2,takahashi3,yoshimura1,yang1,waki1}. Solid line shows a theoretical relation between ${T_{\rm C}}/{T_0}$ and ${p_{\rm eff}}/p_{\rm s}$ in the Takahashi's spin fluctuation theory\cite{takahashi1,takahashi2,takahashi3}.}

\end{figure} 
 \begin{table}[]
\caption{\label{tab:table1}%
Basic magnetic and spin fluctuation parameters for URhSi, and uranium ferromagnetic superconductors URhGe\cite{tateiwa1,tateiwa2}, UCoGe\cite{huy,nksato1} and UGe$_2$\cite{tateiwa1,tateiwa2}. }
\begin{ruledtabular}
\begin{tabular}{ccccccc}
\textrm{}&
\textrm{$T{_{\rm C}}$}&
\textrm{$p_{\rm eff}$}&
\textrm{$p_{\rm s}$}&
\textrm{$T_0$ }&
\textrm{$T_{\rm A}$}&
\textrm{${T_{\rm C}}/T_0$}\\
\textrm{}&
\textrm{(K)}&
\textrm{(${\mu}_{\rm B}$/U)}&
\textrm{(${\mu}_{\rm B}$/U)}&
\textrm{(K)}&
\textrm{(K)}&
\textrm{}\\
\colrule
URhSi&10.5 &2.94 &0.571 &64.5  &354&0.163\\ 
URhGe&9.47   &1.75 & 0.407&78.4 &568&0.121 \\ 
UCoGe&2.4   &1.93 &0.039 &362 &5.92 $\times$ 10$^3$&0.0065 \\ 
UGe$_2$&52.6   &3.00 &1.41  &92.2  &442 &0.571 \\ 
\end{tabular}
\end{ruledtabular}
 \end{table}

 Next, we compare the dynamical magnetic property in URhSi with those in the uranium ferromagnetic superconductors UGe$_2$, URhGe, and UCoGe. Recently, we have studied the applicability of Takahashi's spin fluctuation theory to the actinide $5f$ systems\cite{tateiwa2,takahashi1,takahashi2,takahashi3}. We analyzed the magnetic data of 80 actinide ferromagnets and determined spin fluctuation parameters $T_0$ and $T_{\rm A}$: the widths of the spin fluctuation spectrum in the energy and momentum spaces, respectively. Figure 2 shows the plot of ${p_{\rm eff}}/{p_{\rm s}}$ and ${T_{\rm C}}/{T_{\rm 0}}$ (the generalized Rhodes-Wohlfarth plot) for the actinide ferromagnets, and the $3d$ metals and their intermetallic ferromagnetic compounds\cite{tateiwa2}. The data for uranium, neptunium, and plutonium compounds, and the $3d$ systems are plotted as closed circles, squares, triangles, and anti-triangles, respectively. The data for URhSi, URhGe, UCoGe, and UGe$_2$ are highlighted. A solid line represents a theoretical relation between ${T_{\rm C}}/{T_0}$ and ${p_{\rm eff}}/p_{\rm s}$ in the Takahashi's spin fluctuation theory. The data for UCoGe and the $3d$ systems are cited from the literature\cite{nksato1,takahashi1,takahashi2,takahashi3,yoshimura1,yang1,waki1}. The parameters of the other actinide compounds were determined by us\cite{tateiwa2}. The data of the actinide ferromagnets follow the theoretical relation for ${T_{\rm C}}/{T_{\rm 0}}$ $<$ 1.0. This suggests the applicability of the theory to most of the actinide ferromagnets. Several data points deviate from the relation near ${T_{\rm C}}/{T_{\rm 0}}{\,}={\,}1$, which may be due to some effects arising from the localized character of the $5f$ electrons not included in the theory. In the spin fluctuation theory, the degree of the itinerancy of magnetic electrons can be discussed from the parameter ${T_{\rm C}}/{T_{\rm 0}}$ \cite{takahashi1,takahashi2,takahashi3}. The strong itinerant character of the magnetic electrons is suggested at ${T_{\rm C}}/{T_{\rm 0}}{\,}{\ll}{\,}1$ and a relation ${T_{\rm C}}/{T_{\rm 0}}{\,}={\,}1$ indicates the local moment ferromagnetism. The value of ${T_{\rm C}}/{T_{\rm 0}}$ = 0.571 for UGe$_2$ suggests that it is located comparably close to the local moment system. Meanwhile, the small values of ${T_{\rm C}}/{T_{\rm 0}}$ = 0.0065 and $p_{\rm s}$ for UCoGe suggest the weak ferromagnetism, similar to those in Y(Co$_{1-x}$Al$_x$)$_2$\cite{yoshimura1} and Ni$_3$Al$_{1-x}$Ga$_x$\cite{yang1}. URhSi and URhGe are located in an intermediate region between the two limiting cases. The two uranium ferromagnets share several similarities in terms of the basic physical and the dynamical magnetic properties.

\section{EXPERIMENT and ANALYSIS}
A single-crystal sample of URhSi was grown by Czochralski pulling in a tetra arc furnace. The value of the residual resistivity ratio (RRR) ( = ${{\rho}_{\rm  RT}}$/${{\rho}_0}$) is about 2.5. Here, ${{\rho}_{\rm  RT}}$ and ${{\rho}_0}$ represent the resistivity value at room temperature and the residual resistivity at low temperatures, respectively. Impurities or misoriented grains were not detected in the x-ray diffraction experiment on the single-crystal sample for this study. Magnetization was measured in a commercial superconducting quantum interference (SQUID) magnetometer (MPMS, Quantum Design). The internal magnetic field ${{\mu}_0}H$ was obtained by subtracting the demagnetization field $DM$ from the applied magnetic field ${{\mu}_0}H_{\rm ext}$: ${{\mu}_0}H$ = ${{\mu}_0}H_{\rm ext}$ - $DM$. The demagnetizing factor $D$ (= 0.22) was calculated from the macroscopic dimensions of the sample. We determine the critical exponents using a modified Arrott plot, critical isotherm analysis, a Kouvel-Fisher plot, and scaling analysis.

  \begin{figure}[t]
\includegraphics[width=8 cm]{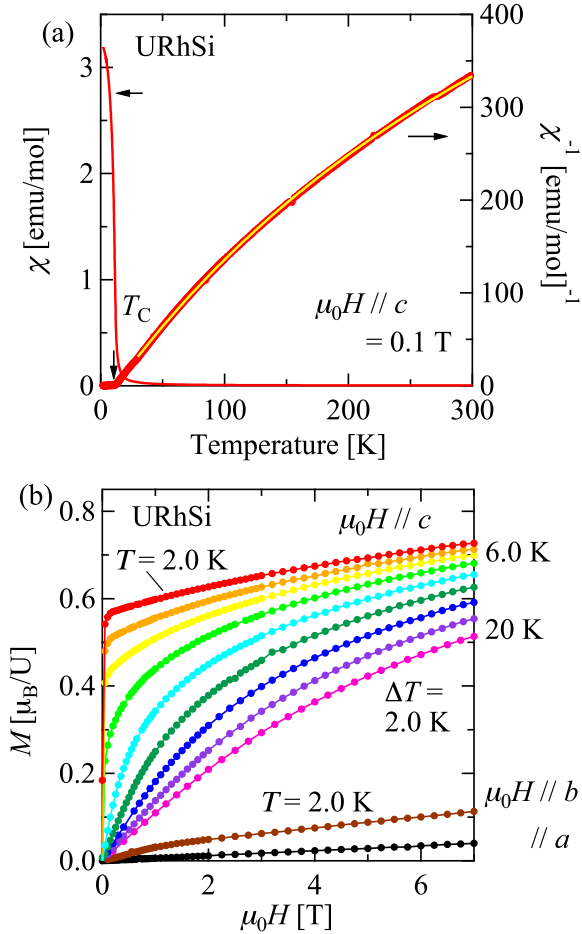}
\caption{\label{fig:epsart}(a)Temperature dependencies of the magnetic susceptibility ${\chi}$ and its inverse $1/{\chi}$ in a magnetic field of 0.1 T applied along the magnetic easy $c$ axis in URhSi. Solid line represents the result of the fit to the inverse of the magnetic susceptibility $1/{\chi}$ using a modified Curie-Weiss law. (b) Magnetic field dependencies of the magnetization at several temperatures in magnetic field applied along the $c$ axis, and the magnetization at 2.0 K in fields along the magnetic hard $b$ and $a$ axes in URhSi.}
\end{figure}

\section{RESULTS}
 We show the temperature dependencies of the magnetic susceptibility ${\chi}$ and its inverse $1/{\chi}$ in a magnetic field of 0.1 T applied along the magnetic easy $c$ axis of URhSi in Fig. 3(a). The magnetic susceptibility ${\chi}$ was analyzed using a modified Curie-Weiss law ${\chi}$ = ${C/(T-{\theta})}+ {{\chi}_0}$ shown as solid line. Here, $C$ and $\theta$ are the Curie constant and the paramagnetic Curie temperature, respectively. $ {{\chi}_0}$ is the temperature-independent component of the magnetic susceptibility from the density of states at the Fermi energy from other than the $5f$ electrons. The effective magnetic moment $p_{\rm eff}$ is determined as $p_{\rm eff}$ = 2.94 ${\mu}_{\rm B}$/U per uranium atom from $C$ = ${N_{\rm A}}{{\mu}_{\rm B}^2}{p_{\rm eff}^2}/3{k_{\rm B}}$. Here, $N_{\rm A}$ is the Avogadro constant. The smaller value of $p_{\rm eff}$ than those expected for $5f^2$ (U$^{4+}$, $p_{\rm eff}$ = 3.58 ${\mu}_{\rm B}$/U) and $5f^3$ (U$^{3+}$, $p_{\rm eff}$ = 3.62 ${\mu}_{\rm B}$/U) configurations suggests the itinerant character of the $5f$ electrons in URhSi. We show the magnetic field dependencies of the magnetization at several temperatures in magnetic field applied along the magnetic easy $c$ axis of URhSi in Fig. 3 (b). The spontaneous magnetic moment $p_{\rm s}$ is determined as $p_{\rm s}$ = 0.571 ${\mu}_{\rm B}$/U from the magnetization curve at 2.0 K. The value of $p_{\rm s}$ is consistent with those ($0.50-0.55$ ${\mu}_{\rm B}$/U) determined by the elastic neutron scattering studies\cite{tran,prokes1}. The value is smaller than the magnetic moment $\mu$ [= ${\mu}^{\rm U}$+${\mu}^{\rm Rh}$ = 0.66(2)+0.05(2) = 0.71(4) ${\mu}_{\rm B}$] determined at 2 K with magnetic field of 6 T by the polarized neutron scattering experiment\cite{prokes3}. The reason for this discrepancy is not clear. The magnetization curves in fields along the magnetic hard $a$ and $b$ axes at 2.0 K are also shown in Fig. 3 (b). Clearly, the ferromagnetic ordered state has large magnetic anisotropy, similar to those in the uranium ferromagnetic superconductors UGe$_2$, URhGe, and UCoGe\cite{sakon,hardy,huy2}. This uniaxial magnetic anisotropy is consistent with the collinear ferromagnetic structure with the uranium magnetic moments oriented along the $c$ axis determined in the neutron scattering studies\cite{tran,prokes1}. 
   
 There are differences in the magnetization curves between the present and the previous studies\cite{prokes1,prokes2}. In the previous studies, the value of $p_{\rm s}$ is less than 0.5 ${\mu}_{\rm B}$/U in the magnetization curve along the $c$ axis and the spontaneous magnetic moment above 0.1 ${\mu}_{\rm B}$/U occurs in magnetic fields applied along the magnetic hard $a$ and $b$ axes. The magnetization along the $a$ axis is slightly larger than that along the $b$ axis at 2.0 K\cite{prokes2,prokes1}. These features are not consistent with the present data shown in Fig. 3 (b). The magnetization curves in the previous studies suggest the tilt of the magnetic moment from the $c$ axis to the $a$-$b$ plane. However, this is inconsistent with the simple ferromagnetic structure with the magnetic moments aligned along the $c$ axis determined by the neutron scattering experiments\cite{tran,prokes1}. The authors of Ref. 22 proposed several possible reasons for this discrepancy such as the existence of grains in their single-crystal sample. It seems that a final conclusion has not been made. We stress that the magnetization curves in Fig. 3 (b) are consistent with the magnetic structure determined in the neutron scattering studies\cite{tran,prokes1}.

 \begin{figure}[t]
\includegraphics[width=7.5cm]{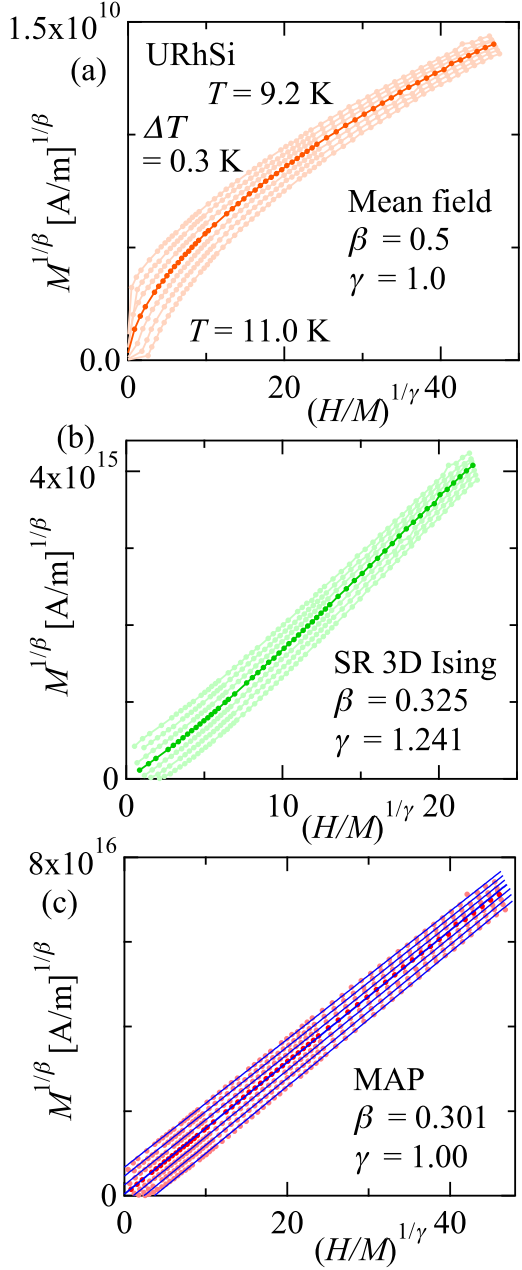}
\caption{\label{fig:epsart}Magnetization isotherms in the forms of $M^{1/{\beta}}$ vs. $(H/M)^{1/{\gamma}}$ in the temperature range 9.2 K $\le$ $T$ $\le$ 11.0 K, with (a) the mean-field theory, (b) the short-range (SR) 3D-Ising model, and (c) the modified Arrott plot (MAP) with ${\beta}$ = 0.301 and ${\gamma}$ = 1.00 in URhSi. Bold circles indicate the isotherms at 10.1 K. Solid lines in (c) show fits to the data with Eq. (6).}
\end{figure}

In the mean-field theory, the free energy of a ferromagnet in the vicinity of $T_{\rm C}$ can be expressed as a power-series expansion in the order parameter $M$:
  \begin{eqnarray}
F(M)&= F(0)+&{1\over 2}aM^2+{1\over 4}bM^4 +....-HM. 
  \end{eqnarray}
 The following equation of state is derived from the equilibrium condition by minimizing the free energy ${\partial F}(M)$/${\partial M}$ = 0:
      \begin{eqnarray}
H&=&aM+bM^3.
  \end{eqnarray}
  
  \begin{figure}[t]
\includegraphics[width=8cm]{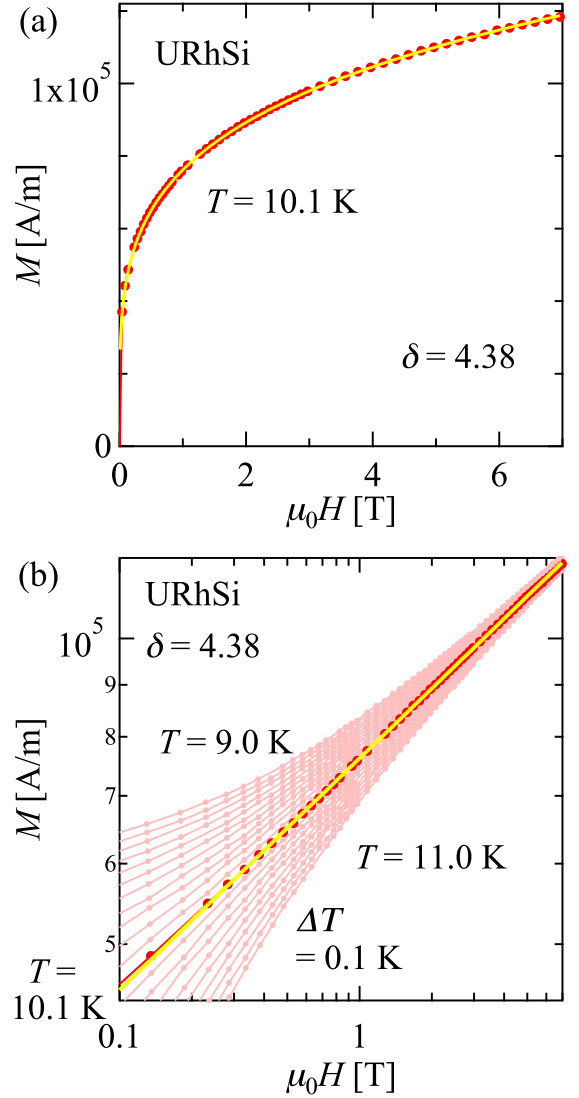}
\caption{\label{fig:epsart}Magnetic field dependencies of the magnetization (a) at 10.1 K and (b) from 9.0 to 11.0 K in URhSi. Bold circles indicate the critical isotherm data at 10.1 K. Solid lines represent fits to the critical isotherm with Eq. (5).}
\end{figure} 
 
 The mean-field theory fails in the asymptotic critical region whose extent can be estimated by the Ginzburg criterion\cite{ginzburg2}. The correlation length $\xi$ = ${\xi}_0$ $|1-T/{T_{\rm C}}|^{-{\nu}}$ diverges in the critical region, which leads to universal scaling laws for the spontaneous magnetization $M_{\rm s}$, the initial susceptibility ${\chi}$, and the magnetization at $T_{\rm C}$. Here, ${{\nu}}$ is the critical exponent. From the scaling hypothesis, the spontaneous magnetization $M{_{\rm s}}(T)$ below $T_{\rm C}$, the inverse of the initial magnetic susceptibility ${\chi}(T)$ below and above $T_{\rm C}$, and the magnetization $M({{\mu}_0}H)$ at $T{_{\rm C}}$ are characterized a set of critical exponents as follows\cite{privman}:   
      \begin{eqnarray}
 M{_{\rm s}}(T) &{\propto}& |t|{^{\beta}}  {\;}  {\;} {\;}{\;}{\;} (T < T{_{\rm C}}),\\ 
{{\chi}}(T){^{-1}}&{\propto}&  {|t|}^{{{\gamma}}'}  {\;}  {\;} (T<T{_{\rm C}}), {\;} {|t|}^{{{\gamma}}} {\;}{\;}(T{_{\rm C}}< T),\\ 
 {M}({{\mu}_0}H)& {\propto} & ({{\mu}_0}H)^{1/{\delta}} {\;} {\;}(T = T{_{\rm C}}).
   \end{eqnarray}        
 Here, $t$ is the reduced temperature $t$ = $1-{T}/{T_{\rm C}}$. $\beta$, $\gamma$, ${\gamma}'$, and $\delta$ are the critical exponents. 

 \begin{figure}[]
\includegraphics[width=8.0cm]{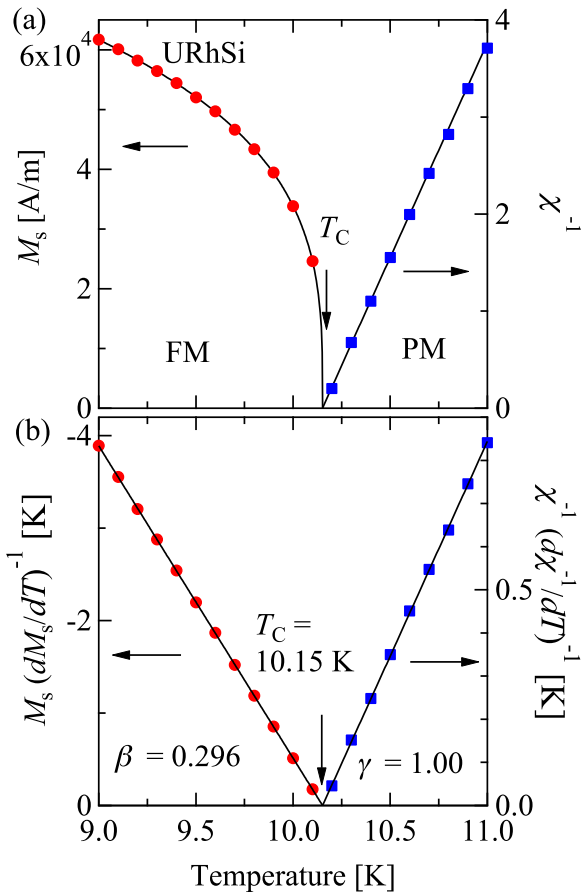}
\caption{\label{fig:epsart}(a) Temperature dependencies of the spontaneous magnetization $M{_{\rm s}}(T)$ below $T_{\rm C}$ (left) and the inverse of the initial magnetic susceptibility ${\chi}^{-1}$ above $T_{\rm C}$ (right) determined from the modified Arrott plot. (b)Kouvel-Fisher plots of $M{_{\rm s}}(T)[dM{_{\rm s}}(T)/dT]{^{-1}}$ (left) and $ {{\chi}}{^{-1}}(T)[d {{\chi}}{^{-1}}(T)/dT]{^{-1}}$ (right) in URhSi.}
\end{figure} 
         \begin{figure}[]
\includegraphics[width=8.0cm]{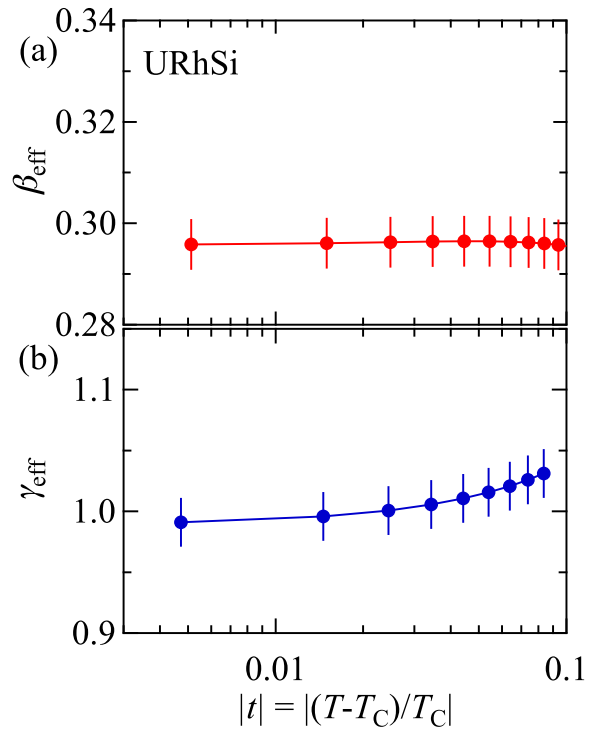}
\caption{\label{fig:epsart}Effective exponents (a) ${\beta}_{\rm eff}$ for the spontaneous magnetization $M{_{\rm s}}(T)$ below $T_{\rm C}$ and (b) ${\gamma}_{\rm eff}$ for the magnetic susceptibility ${\chi}$ above $T_{\rm C}$ as a function of the reduced temperature $|t|$ [=$|({T}-{T_{\rm C}})/{T_{\rm C}}|$] in URhSi.}
\end{figure} 

    \begin{figure}[t]
\includegraphics[width=8cm]{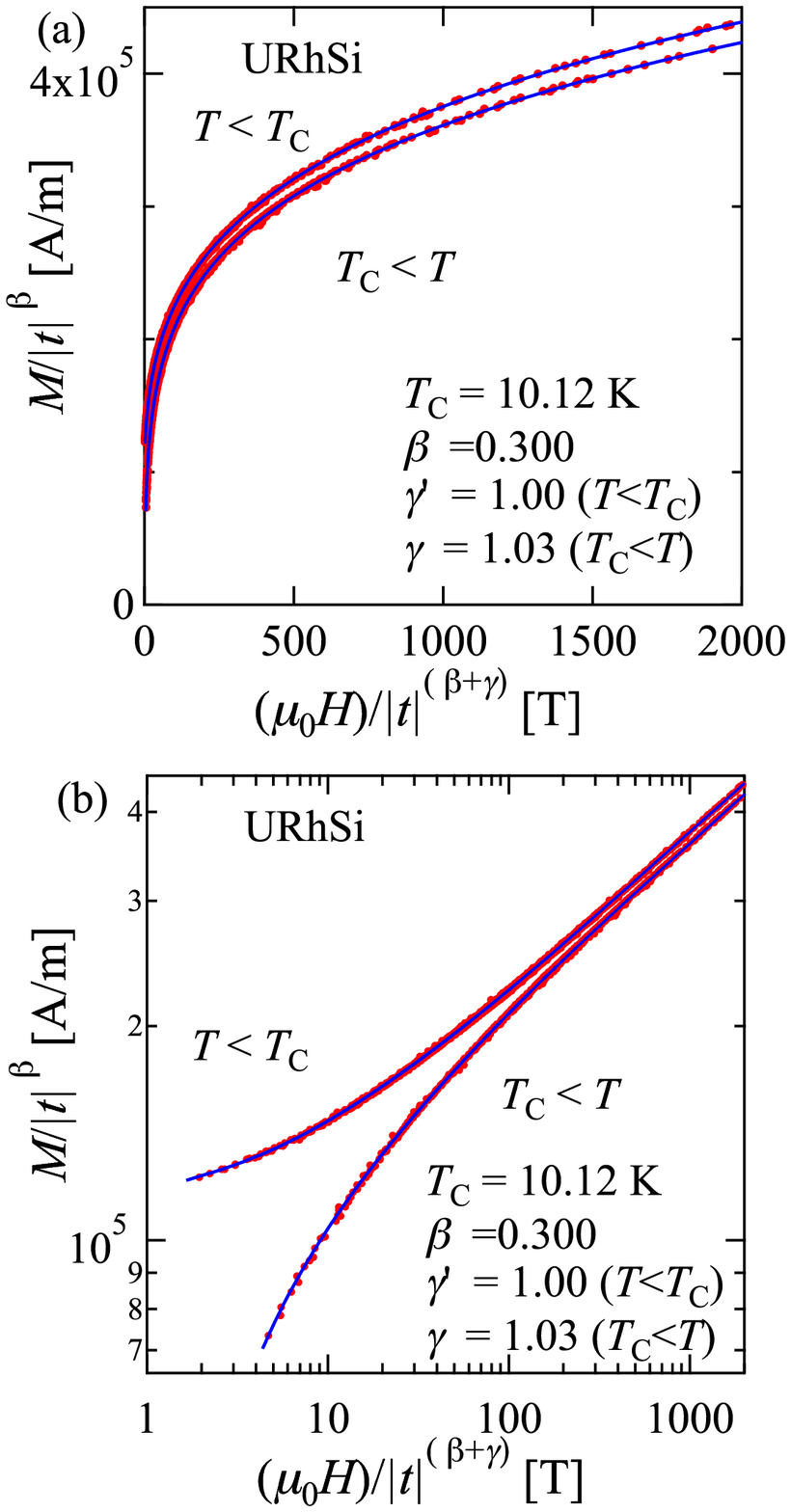}
\caption{\label{fig:epsart}Renormalized magnetization $m$ ($\equiv$ ${|t|^{-{\beta}}}{M({{\mu}_0}H, t)}$) of URhSi as a function of renormalized field $h$ [$\equiv$ ${H}{|t|^{-({\beta}+{\gamma})}}$] following Eq. (11) below and above $T_{\rm C}$ with $T{_{\rm C}}$, $\beta$, ${\gamma}$, and ${\gamma}'$ values mentioned in the main text. Solid lines represent best-fit polynomials. The magnetization data in the temperature range $t{\,}={\,}|({T}-{T_{\rm C}})/{T_{\rm C}}|{\,}< 0.1$ are plotted.}
\end{figure}

 Usually, the Arrott plots technique has been used to determine the phase transition temperature $T{_{\rm C}}$. In the mean-field theory, isotherms plotted in the form of $M^2$ vs. $H/M$ should be a series of parallel straight lines and the isotherm at $T_{\rm C}$ should pass through the origin\cite{privman}. The critical exponents with $\beta$ = 0.5, $\gamma$ = 1.0, and $\delta$ = 3.0 in the mean-field theory are assumed in the Arrott plot. 
 
  Figures 4 (a) and (b) show the magnetization isotherms in the forms of $M^{1/{\beta}}$ vs. $(H/M)^{1/{\gamma}}$ with (a) the mean-field theory ($\beta$ = 0.5 and $\gamma$ = 1.0) and (b) the 3D-Ising model with short-range (SR) exchange interactions ($\beta$ = 0.325 and $\gamma$ = 1.241), respectively. The isotherms do not form straight lines in the two plots. Therefore, the Arrott-Noakes equation of state has been used to re-analyze the magnetization isotherms\cite{arrott}. The following equation should hold in the asymptotic critical region. 

  \begin{eqnarray}
  &&(H/M){^{1/{\gamma}}} = (T-{T_{\rm C}})/{T_1} + (M/{M_{1}})^{1/{\beta}}
   \end{eqnarray}

Here, $T_1$ and $M_1$ are material constants. The data for URhSi are plotted in the form of $M^{1/{\beta}}$ vs. $(H/M)^{1/{\gamma}}$ in the modified Arrott plots. The isotherms exhibit a linear behavior when the appropriate values of $T_{\rm C}$, $\beta$, and $\gamma$ are chosen as shown in Fig. 4 (c). The values are determined as $T_{\rm C}$ = 10.12 $\pm$ 0.02 K, $\beta$ = 0.301 $\pm$ 0.002, and ${\gamma}$ = 1.00 $\pm$ 0.04 from a best fit of Eq. (6) to the data for 9.2 K $\le$ $T$ $\le$ 11.0 K and 0.1 T $\le$ ${{\mu}_0}H$ $\le$ 7.0 T in URhSi.

 The third critical exponent $\delta$ is determined as $\delta$ = 4.38 $\pm$ 0.04 for URhSi from fits to the critical isotherm at 10.1 K with Eq. (5) as shown in Figure 5. The value is lower than that in the 3D Ising model with short-range exchange interactions ($\delta$ = 4.82). The exponents $\beta$, $\gamma$, and $\delta$ should fulfill the Widom scaling law ${\delta}$ = 1+${\gamma}/{\beta}$\cite{widom}. We estimate the value of ${\delta}$ as 4.32 $\pm$ 0.10 from the $\beta$ and $\gamma$ values determined in the modified Arrott plots using the law. This value is consistent with that determined from the critical isotherm. 

The data are analyzed using the Kouvel-Fisher (KF) method by which the critical exponents $\beta$ and $\gamma$ can be determined more accurately\cite{kouvel}. At first, we determine the temperature dependencies of the spontaneous magnetization $M{_{\rm s}}$ and the initial magnetic susceptibility ${\chi}$ from the modified Arrott plots as follows. The fitted straight lines in the plots intersect with the vertical axis at $M^{1/{\beta}}$ = $M{_{\rm s}}^{1/{\beta}}$ for $T$ $<$ $T_{\rm C}$ and with the transverse axis at $(H/M)^{1/{\gamma}}$ = $(1/{\chi})^{1/{\gamma}}$ for $T_{\rm C}$ $<$ $T$\cite{seeger}. Next, the temperature dependencies of $M{_{\rm s}}$ and ${\chi}^{-1}(T)$ are obtained by inserting the values of the exponents $\beta$ and $\gamma$ determined in the modified Arrott plots. Figure 6 (a) shows the temperature dependencies of $M{_{\rm s}}$ and ${\chi}^{-1}(T)$ in URhSi. Solid lines show the fits to the data using Eqs. (3) and (4) for $M{_{\rm s}}(T)$ and ${\chi}^{-1}(T)$, respectively. In the KF method, temperature-dependent exponents ${\beta}(T)$ and ${\gamma}(T)$ are defined as follows:
 
       \begin{eqnarray}
 M{_{\rm s}}(T)[dM{_{\rm s}}(T)/dT]{^{-1}} &=& (T-T{_{\rm C}}^{-})/{\beta}(T),\\ 
  {{\chi}}{^{-1}}(T)[d {{\chi}}{^{-1}}(T)/dT]{^{-1}} &=& (T-T{_{\rm C}}^{+})/{\gamma}.(T)
   \end{eqnarray}

  Equations (7) and (8) can be obtained from Eq. (6) in the limit $H$ $\rightarrow$ 0 for $T$ $<$ and $>$ $T_{\rm C}$, respectively. We determine the values of $\beta$ and $\gamma$ from the slope of $M{_{\rm s}}(T)[dM{_{\rm s}}(T)/dT]{^{-1}}$ and $ {{\chi}}{^{-1}}(T)[d {{\chi}}{^{-1}}(T)/dT]{^{-1}}$-plots, respectively, at $T_{\rm C}$ as shown in Fig. 6 (b).  Note that the quantities ${\beta}(T)$ and ${\gamma}(T)$ in the limit $T$ $\rightarrow$ $T_{\rm C}$ correspond to the critical exponents $\beta$ and $\gamma$, respectively. Solid lines in Fig. 6 (b) show the fits to the data using Eqs. (7) and (8). The determined values of the exponents $\beta$ and $\gamma$ are $\beta$ = 0.296 $\pm$ 0.002 and $\gamma$ = 1.00 $\pm$ 0.02 with $T_{\rm C}$ = ($T{_{\rm C}}^{+}$ + $T{_{\rm C}}^{-}$)/2 = 10.15 $\pm$ 0.01 K. The determined exponents are consistent with those determined in the modified Arrott plot. 

If there are various competing interactions or disorders, crossover phenomena could occur in the critical exponents on approaching $T_{\rm C}$ as observed in Ni$_3$Al\cite{semwal}. The convergence of the critical exponents should be checked. Effective exponents ${\beta}_{\rm eff}$ and ${\gamma}_{\rm eff}$ are useful to examine this possibility.

   \begin{eqnarray}
{{\beta}_{\rm eff}} (t) =  d[{\rm ln}M{_{\rm s}}(t)]/d({{\rm ln}{t}}),\\
{{\gamma}_{\rm eff}} (t) =  d[{\rm ln}{{\chi}^{-1}}(t)]/d({{\rm ln}{t}}).
  \end{eqnarray}

We show the effective exponents ${\beta}_{\rm eff}$ and ${\gamma}_{\rm eff}$ as a function of the reduced temperature $t$ in Figs 7 (a) and (b), respectively. A monotonic $|t|$ dependence is observed in both ${\beta}_{\rm eff}$ and ${\gamma}_{\rm eff}$ for ${|t|}{\,}{\geq}$ 5.12${\times}{10^{-3}}$ and 4.73${\times}{10^{-3}}$, respectively. The crossover phenomenon in the critical behavior between two universality classes can be ruled out.

     \begin{table*}[]
\caption{\label{tab:table1}%
Comparison of critical exponents $\beta$, $\gamma$, ${\gamma}'$, and $\delta$ of various theoretical models\cite{privman,fisher0,guillou} with those in URhSi, uranium ferromagnetic superconductors UGe$_2$\cite{tateiwa1}, URhGe\cite{tateiwa1} and UCoGe\cite{huy}, and uranium ferromagnets  UIr\cite{knafo} and U(Co$_{0.98}$Os$_{0.02}$)Al\cite{maeda}. Abbreviations: RG-${\phi}^4$, renormalization group ${\phi}^4$ field theory; SR, short-range; LR, long-range.}
\begin{ruledtabular}
\begin{tabular}{ccccccccc}
\textrm{}&
\textrm{Method}&
\textrm{$T{_{\rm C}}$(K)}&
\textrm{$\beta$ }&
\textrm{${\gamma}'$}&
\textrm{${\gamma}$}&
\textrm{$\delta$}&
\textrm{Reference}&\\
\textrm{}&
\textrm{}&
\textrm{}&
\textrm{}&
\textrm{($T<T{_{\rm C}}$)}&
\textrm{($T{_{\rm C}}<T$)}&
\textrm{}&
\textrm{}&\\
\colrule
(Theory)&&&&&&&&\\ 
Mean-field &&&0.5&\multicolumn{2}{c}{1.0}&3.0&&\\ 
SR exchange: $J(r){\,}{\sim}{\,}e^{-r/b}$ &&&&&&&&\\ 
$d$ =  2, $n$ =1 &Onsager solution&&0.125&\multicolumn{2}{c}{1.75}&15.0&\cite{privman,fisher0}&\\ 
$d$ =  3, $n$ =1 &RG-${\phi}^4$&&0.325&\multicolumn{2}{c}{1.241}&4.82&\cite{guillou}&\\ 
$d$ =  3, $n$ =2 &RG-${\phi}^4$&&0.346&\multicolumn{2}{c}{1.316}&4.81&\cite{guillou}&\\ 
$d$ =  3, $n$ =3 &RG-${\phi}^4$&&0.365&\multicolumn{2}{c}{1.386}&4.80&\cite{guillou}&\\ 
\colrule
URhSi &&&&&&&This work&\\
&Modified Arrott &10.12 $\pm$ 0.02  &0.301 $\pm$ 0.002 &  \multicolumn{2}{c}{1.00 $\pm$ 0.04} &&&\\ 
&Kouvel-Fisher &10.15  $\pm$ 0.01  &0.296 $\pm$ 0.002 &  \multicolumn{2}{c}{1.00 $\pm$ 0.02}  &&&\\
&Scaling &10.12 $\pm$ 0.02  &0.300 $\pm$ 0.002& 1.00 $\pm$ 0.02 & 1.03 $\pm$ 0.02 &&& \\
&Critical isotherm &  & & && 4.38 $\pm$ 0.04 && \\
UGe$_2$&&&&&&&\cite{tateiwa1}&\\
&Modified Arrott &52.6 $\pm$ 0.1  &0.334 $\pm$ 0.002 &  \multicolumn{2}{c}{1.05 $\pm$ 0.05} &&&\\ 
&Kouvel-Fisher &52.60  $\pm$ 0.02  &0.331 $\pm$ 0.002 &  \multicolumn{2}{c}{1.03 $\pm$ 0.02}  &&&\\
&Scaling &52.79  $\pm$ 0.02  &0.329 $\pm$ 0.002& 1.00 $\pm$ 0.02 & 1.02 $\pm$ 0.02 &&& \\
&Critical isotherm &  & & && 4.16 $\pm$ 0.02 && \\
URhGe&&&&&&&\cite{tateiwa1}&\\
&Modified Arrott &9.44 $\pm$ 0.02  &0.303 $\pm$ 0.002 &  \multicolumn{2}{c}{1.02 $\pm$ 0.03} &&&\\ 
&Kouvel-Fisher &9.47  $\pm$ 0.01  &0.303 $\pm$ 0.002 &  \multicolumn{2}{c}{1.01 $\pm$ 0.02}  &&&\\
&Scaling &9.47  $\pm$ 0.01  &0.302 $\pm$ 0.001& 1.00 $\pm$ 0.01 & 1.02 $\pm$ 0.01 &&& \\
&Critical isotherm &  & & && 4.41 $\pm$ 0.02 && \\
UCoGe& &2.5  &  \multicolumn{4}{c}{$\sim$ mean-field type $\sim$ } &\cite{huy}&\\ 
UIr&&&&&&&\cite{knafo}&\\ 
&Modified Arrott &45.15  &0.355(50) &  \multicolumn{2}{c}{1.07(10)} &&&\\ 
&Critical isotherm &  & & && 4.01(5) && \\
U(Co$_{0.98}$Os$_{0.02}$)Al& & &&   &&&\cite{maeda}&\\ 
&Modified Arrott &25  &0.33 &  \multicolumn{2}{c}{1.0} &&&\\ 
&Critical isotherm &  & & && 4.18 && \\
\end{tabular}
\end{ruledtabular}
 \end{table*}

 It is necessary to examine the possibility of the strongly asymmetric critical region or the change of the universality class across $T_{\rm C}$. The values of ${\gamma}$' for $T<T{_{\rm C}}$ and $\gamma$ for $T{_{\rm C}}<T$ can be determined separately with the scaling theory where a reduced equation of state close to $T_{\rm C}$ is expressed as follows\cite{privman}: 
    \begin{eqnarray}
 m = f{_{\pm}}{(h)}. 
 \end{eqnarray}
Here, $f_{+}$ for $T{_{\rm C}} < T$ and $f_{-}$ for $T < T{_{\rm C}}$ are regular analytical functions. The renormalized magnetization $m$ and field $h$ are defined as $m$ $\equiv$ ${|t|^{-{\beta}}}{M({{\mu}_0}H, t)}$ and $h$ $\equiv$ ${H}{|t|^{-({\beta}+{\gamma})}}$, respectively. This relation implies that the data of $m$ versus $h$ with the correct values of $\beta$, ${\gamma}$', $\gamma$, and $t$ fall on two universal curves, one for $T < T{_{\rm C}}$ and the other for $T{_{\rm C}} < T$. We show the renormalized magnetization $m$ as a function of the renormalized field $h$ below and above $T_{\rm C}$ in Figs. 8 (a) and (b). The magnetization data in the temperature range $t{\,}={\,}|({T}-{T_{\rm C}})/{T_{\rm C}}|{\,}< 0.1$ are plotted. All data points collapse onto two independent curves. The values of $T_{\rm C}$ and the critical exponents are determined as $T_{\rm C}$ = 10.12 $\pm$ 0.02 K, $\beta$ = 0.300 $\pm$ 0.002, ${\gamma}'$ = 1.00 $\pm$ 0.02 for $T$ $<$ $T_{\rm C}$, and $\gamma$ = 1.03 $\pm$ 0.02 for $T_{\rm C}$ $<$ $T$ in URhSi. The scaling analysis suggests that the set of the critical exponents are the same below and above $T_{\rm C}$ in URhSi. We can rule out the strongly asymmetric critical region and the change of the universality class across $T_{\rm C}$.

Table III shows the critical exponents $\beta$, $\gamma$, ${\gamma}'$, and $\delta$ in various theoretical models\cite{privman,fisher0,guillou} and those in URhSi. We also show the exponents in the uranium ferromagnetic superconductors UGe$_2$\cite{tateiwa1}, URhGe\cite{tateiwa1}, and UCoGe\cite{huy}, and uranium ferromagnets UIr with $T_{\rm C}$ of 46 K\cite{knafo,galatanu} and U(Co$_{0.98}$Os$_{0.02}$)Al with $T_{\rm C}$ of 25 K\cite{maeda,andreev}. The strong uniaxial magnetization in the ferromagnetic state of URhSi suggests the universality class of the 3D Ising model. However, the obtained critical exponents in URhSi are different from those of the 3D Heisenberg ($d$ =  3, $n$ =3), 3D XY ($d$ =  3, $n$ =2), 3D Ising ($d$ =  3, $n$ =1), and 2D Ising ($d$ =  2, $n$ =1) models where magnetic moments are interacted via short-range (SR) exchange interactions of a form $J(r){\,}{\sim}{\,}e^{-r/b}$. Here, $b$ is the correlation length. The value of $\beta$ in URhSi is close to those in the 3D models but the $\gamma$ value is close to unity, expected one in the mean-field theory. While the magnetization $M_{\rm s}$ shows the critical behavior around $T{_{\rm C}}$, the magnetic susceptibility $\chi$ follows the mean-field theory in URhSi. The universality class of the ferromagnetic transition in URhSi may not belong to any known universality class.

 The obtained critical exponents in URhSi are similar to those in UGe$_2$, URhGe, UIr, and U(Co$_{0.98}$Os$_{0.02}$)Al where the ferromagnetic state has strong uniaxial magnetic anisotropy. The universality class of the ferromagnetic transitions in URhSi may belong to the same one of the uranium ferromagnets. Note that we previously reported the unconventional critical scaling of magnetization in UGe$_2$ and URhGe\cite{tateiwa1}. The values of the exponent $\beta$ slightly differ, depending on each ferromagnet. Meanwhile, the $\gamma$ values of the uranium ferromagnets are close to unity. This almost mean-field behavior of the magnetic susceptibility ${\chi}$ may be a characteristic feature of the unconventional critical behavior of the magnetization in the uranium ferromagnets. The ferromagnetic correlation in the uranium ferromagnets may be different from that in the 3D Ising system.  This unusual critical behavior of the magnetization may be inherent in the ferromagnetism of $5f$ electrons where the superconductivity could appear.

 We discuss the extent of the asymptotic critical region where magnetic data for the determination of the critical exponents should be collected. The extent of the region can be estimated by the Ginzburg criterion\cite{ginzburg2,chaikin,yelland}:
   \begin{eqnarray}
  &&{\Delta}{T_{\rm G}}/{T_{\rm C}} = {k_{\rm B}^2}/[32{{\pi}^2}{({\Delta}C)^2}{{\xi_0}^6}].
   \end{eqnarray}
 Here, ${\Delta}C$ is the jump of the specific heat at $T_{\rm C}$ and ${\xi}_0$ is the correlation length $\xi$ ($1/{{\xi}^{2}}$ = $1/{{\xi}_0^{2}}|1-T/{T_{\rm C}}|$) at 2${T_{\rm C}}$.  It is possible to estimate the temperature region where the mean-field theory fails by the Ginzburg criterion. Previously, we determined the value of ${\Delta}{T_{\rm G}}$ as $\sim$ 100 K for UGe$_2$ in Ref. 18 using the neutron scattering and specific heat data\cite{huxley1,huxley3} and concluded that the data used for the determination of the critical exponents in UGe$_2$ were collected inside the asymptotic critical region. The large value of ${\Delta}{T_{\rm G}}$ may be due to experimental errors in the values of ${\Delta}C$ and ${\xi}_0$. ${\Delta}{T_{\rm G}}$ is very sensitive to ${\xi}_0$. Unfortunately, it is impossible to estimate ${\Delta}{T_{\rm G}}$ for URhSi since there has been no report for the correlation length ${\xi}_0$. The critical exponents in URhSi are determined using the data in the temperature region from 9.0 K to 11.0 K (0 $<$ $t$ $<$ 0.1). The present analyses suggest that this temperature region is inside the asymptotic critical region. The $T$-linear dependence of the magnetic susceptibility ${\chi}^{-1}$ does not indicate that the analysis is done using the data taken outside the asymptotic critical region. We also rule out the possibility of the strongly asymmetric critical region or the change of the universality class across $T_{\rm C}$ as mentioned before.

The mean-field behavior of the magnetization in UCoGe is briefly discussed\cite{huy}. As shown in Table II, the values of the spontaneous magnetic moment ($p_{\rm s}$ = 0.039 ${\mu}_{\rm B}$/U) and the parameter ${T_{\rm C}}/{T_{\rm 0}}$ ( = 0.0065) are very small compared with those in UGe$_2$ and URhGe. These results suggest the strong itinerant characters of the $5f$ electrons. We previously estimated the value of ${\Delta}{T_{\rm G}}$ as less than 1 mK using the specific heat and the neutron data\cite{huy,tateiwa1,stock}. This value suggests a very narrow asymptotic critical region. The strong itinerant character of the $5f$ electrons masks the critical behavior in UCoGe. The mean-field behavior of the magnetization is expected to appear since most of the magnetic data around $T_{\rm C}$ might be collected outside the very narrow asymptotic critical region.

 We estimate the critical exponent $\alpha$ for the specific heat [$C(T){\,}{\propto}{\,}|t|{^{\alpha}}$] as $\sim$ 0.4 using the Rushbrooke scaling relation ($\alpha$ + 2$\beta$ + $\gamma$ = 2)\cite{rushbrooke}. In the mean field theory (${\alpha}$ = 0), the specific heat does not exhibit divergence at $T_{\rm C}$ and there is no contribution from the magnetic critical fluctuations to the specific heat ($C_{\rm mag}$ = 0) above $T_{\rm C}$. The $\alpha$ value in URhSi suggests the significant contribution to $C_{\rm mag}$ ($>$ 0) from the critical fluctuations above $T_{\rm C}$. This is consistent with a specific heat tail in the temperature range 0 $<$ $t{\,}[= |({T}-{T_{\rm C}})/{T_{\rm C}}|]$ $<$ $\sim$ 0.1 above $T_{\rm C}$\cite{prokes2}. 

The critical exponents in URhSi were previously reported as $\beta$ = 0.36 $\pm$ 0.02 and $\gamma$ = 1.14 $\pm$ 0.06 from the analysis of the magnetization with the scaling theory in Ref. 20. The values of $\beta$ and $\gamma$ are larger than those in the present study. In the magnetization data reported in Refs. 20 and 22, the spontaneous magnetic moment occurs in magnetic fields applied along the magnetic hard $a$ and $b$ axes as mentioned before. The magnetization curves in the previous studies are not compatible with the simple ferromagnetic structure with the magnetic moments oriented along the $c$ axis determined in the neutron scattering studies\cite{tran, prokes1}. Meanwhile, the magnetization curves in this study shown in Fig. 3 (b) are consistent with the magnetic structure. In this study, the critical exponents $\beta$ and $\gamma$ are determined by several different methods: the modified Arrott plot, the Kouvel-Fisher plot, and the scaling analysis. The values of $\beta$, $\gamma$, and $\delta$ satisfy the Widom scaling law (${\delta}$ = 1+${\gamma}/{\beta}$). 
  
\section{DISCUSSIONS}
\subsection{Unconventional critical behavior of magnetization}
 We discuss the unconventional critical behavior of the magnetization in URhSi, UGe$_2$, URhGe, UIr, and U(Co$_{0.98}$Os$_{0.02}$)Al with previous theoretical approaches to critical phenomena.

 (1) The universality class of the magnetic phase transition is affected by the long-range nature of the magnetic exchange interaction. The strength of the magnetic exchange interaction $J(r)$ decreases rapidly with distance in the theoretical models with short-range (SR) interactions. The exchanged interaction can be long-ranged for the itinerant electron system. When the range of the interaction becomes longer, the critical exponents of each universality class are shifted towards those in the mean-field theory. Fischer {\it et al.} analyzed systems with the exchange interaction of a form $J(r)$ $\sim$ $1/r^{d+{\sigma}}$ by a renormalization group approach\cite{fisher1}. Here, $d$ is the dimension of the system and $\sigma$ is the range of exchange interaction. They showed the validity of such a model with long-range interactions for $\sigma$ $<$ 2 and derived a theoretical formula for the exponent $\gamma$ = $\Gamma{\{}\sigma$,$d$,$n{\}}$. Here, $n$ is the dimension of the order parameter and the function $\Gamma$ is given in Ref. 51. Recently, we studied the critical behavior of the magnetization in URhAl around $T_{\rm C}$ = 26.02 K\cite{tateiwa3}. The critical exponents in URhAl were explained with the result of this renormalization group approach for the 2D Ising model coupled with long-range interactions decaying as $J(r)$ $\sim$ $1/r^{2+{\sigma}}$ with ${\sigma}$ = 1.44. We try to reproduce the critical exponents in URhSi, UGe$_2$, URhGe, UIr, and U(Co$_{0.98}$Os$_{0.02}$)Al using the formula for different sets of ${\{}d:n{\}}$ ($d$, $n$ = 1, 2, 3). However, no reasonable solution of ${\sigma}$ is found.
  
(2) Next, we discuss the effect of classical dipole-dipole interaction on the critical phenomenon. The effect on the critical behavior of the magnetization in gadolinium ($T_{\rm C}$ = 292.7 K, $p_{\rm s}$ = 7.12 ${{\mu}_{\rm B}}$/Gd) has been studied\cite{srinath}. This scenario seems not applicable to the uranium ferromagnets since the strength of the effect depends on the square of the spontaneous magnetic moment $p_{\rm s}$\cite{fisher2}. The theoretical values of the critical exponents for the critical phenomena associated with the isotropic or anisotropic dipole-dipole interaction are not consistent with those in URhSi, UGe$_2$, URhGe, UIr, and U(Co$_{0.98}$Os$_{0.02}$)Al\cite{frey1,frey2}. 

(3) Spin fluctuation theories have been developed to explain the finite temperature magnetic properties in itinerant ferromagnets of the $3d$ metals and their intermetallics\cite{moriya1}. For example, the nearly $T$-linear dependence of ${\chi}^{-1}$ above $T_{\rm C}$ observed in the $3d$ electrons systems has been reproduced in numerical calculations based on Moriya's self-consistent renormalization (SCR) theory\cite{moriya2,moriya3} and the Takahashi's spin fluctuation theory\cite{takahashi1,takahashi2,takahashi3}. It is difficult to discuss the exact temperature dependencies of ${\chi}^{-1}$ in the temperature region close to $T_{\rm C}$. This is because the behavior of ${\chi}^{-1}$ near $T_{\rm C}$ depends on the values of parameters in the theories. Calculated ${\chi}^{-1}$-$T$ curves for certain parameter regions are concave upward near $T_{\rm C}$\cite{moriya2,moriya3}. The $T^{4/3}$-dependence of $p_{\rm s}^2$ was derived in the weak coupling limit by the SCR theory\cite{makoshi} and the dependence was roughly reproduced numerically at certain parameter regions in the Takahashi's spin fluctuation theory\cite{takahashi1,takahashi2,takahashi3}. These results are not consistent with the experimentally observed critical exponents of the uranium ferromagnets. Furthermore, we note that the critical exponents are determined in the asymptotic critical region where the spin fluctuations theories cannot be applied to.

(4) We discuss the critical exponents in the uranium ferromagnets from the viewpoint of the local moment magnetism. The  orthorhombic TiNiSi-type crystal structure of URhSi, URhGe, and UCoGe can be regarded as the coupled zigzag chains of the nearest neighbor uranium atoms as mentioned in the introduction. UGe$_2$ and UIr crystalize in the orthorhombic ZrGa$_2$-type (space group $Cmmm$) and the monoclinic PbBi-type (space group $P2_1$) structures, respectively\cite{huxley1,yamamoto}. The crystal structures also can be regarded as the coupled zigzag chains along the crystallographic $a$ axis for UGe$_2$ and the $b$ axis for UIr. The magnetic structures of these ferromagnets could be mapped onto the anisotropic 3D Ising model or the anisotropic next-nearest-neighbor 3D Ising (ANNNI) model. However, the critical exponents of the uranium ferromagnets are not consistent with those for the two models obtained by numerical calculations\cite{yurishchev,murtazev}. 

(5) Recently, Singh, Dutta, and Nandy discussed the unconventional critical behavior of the magnetization in UGe$_2$ and URhGe with a non-local Ginzburg-Landau model focusing on magnetoelastic interactions that give a nonlocal quartic interaction\cite{singh1}. The authors claimed that the calculated critical exponents are comparable with the experimentally observed critical exponents in UGe$_2$, URhGe, and UIr. We hope that the almost mean-field behavior of the magnetic susceptibility $\chi$ in the uranium ferromagnets is completely reproduced.

  It is difficult to explain the critical exponents in the uranium ferromagnets with previous approaches to critical phenomena as discussed in points $1-5$. Here, we introduce several interesting experimental studies on UGe$_2$ and suggest the relevance of the dual nature of the $5f$ electrons between itinerant and localized characters to the critical behaviors of the magnetization in the uranium ferromagnets\cite{yaouanc,sakarya,haslbeck}. The long correlation length of ${\xi}_0$ =48 {\AA} with a magnetic moment of 0.02 ${\mu}_{\rm B}$/U was detected in UGe$_2$ by the Muon spin rotation spectroscopy\cite{yaouanc}. The value of ${\xi}_0$ is more than two times larger than that ($\sim$ 22 {\AA}) determined by the inelastic neutron scattering experiment\cite{huxley3}. A main contribution to the magnetic scattering intensity in the neutron scattering experiment comes from the localized component of the $5f$ electrons in UGe$_2$ since the intensity is proportional to the square of the magnetic moment. The magnetic moment on the uranium site was determined as ${\mu}^{\rm U}$ = $1.45-1.46$ ${{\mu}_{\rm B}}$/U at 6 K by the polarized neutron scattering experiment\cite{kernavanois}. The longer magnetic correlation with the smaller magnetic moment has been attributed to the itinerant component of the $5f$ electrons\cite{yaouanc,sakarya}. Very recently, Haslbeck {\it et al.} have reported the results of the ultrahigh-resolution neutron scattering experiment\cite{haslbeck}. According to the authors, their results suggest the dual nature of spin fluctuations in UGe$_2$; local spin fluctuations described by the 3D Ising universality class and itinerant spin fluctuations. The concept of the duality of the $5f$ electrons has been employed in theoretical models for the superconductivity in the ferromagnetic state of UGe$_2$ and URhGe\cite{troc,hattori2}, and in the antiferromagnetic state of UPd$_2$Al$_3$\cite{sato,thalmeier}. There might be a Hund-type coupling between the itinerant and localized components of the $5f$ electrons. A novel critical phenomenon could appear due to the different nature of the two correlations and the coupling of the two components.

      
In Fig. 2, we show the results of the analyses on the actinide ferromagnets with the Takahashi's spin fluctuation theory\cite{tateiwa2}. The applicability of the theory to actinide $5f$ systems is discussed. Huxley {\it et al.} reported from the inelastic neutron scattering experiment that ${\chi}(q){\Gamma}_q$ remains large for $q{\,}{\rightarrow}{\,}0$ from the data of ${\chi}(q){\Gamma}_q$ measured for $q$ $\geq$ 0.03 {\AA}$^{-1}$\cite{huxley3}. Here, ${\Gamma}_q$ is the relaxation rate for the magnetization density. This non-Landau damping of magnetic excitations suggests that the uniform magnetization density is not a conserved quantity. This fact may raise doubts about the applicability of spin fluctuation theories to the actinide $5f$ electrons systems. Phenomenological and microscopic theories were proposed to explain this non-zero ${\Gamma}(0)$ focusing on the duality of the $5f$ electrons\cite{mineev1,chubukov}. In the recent experiment by Haslbeck {\it et al.}\cite{haslbeck}, the $q$ dependence of ${{\chi}(q)}{\Gamma_q}$ was determined down to $q$ $\sim$ 0.02 {\AA}$^{-1}$, lower than the low limit of $q$ (= 0.03 {\AA}$^{-1}$) in the previous study\cite{huxley3}. ${{\chi}(q)}{\Gamma_q}$ is almost constant for $q^0$ (= 0.038 {\AA}$^{-1}$) $<$ $q$ but it approaches to zero [${\chi}(q){{\Gamma}_q}{\,}{\rightarrow}{\,}0$] for $q{\,}{\rightarrow}{\,}0$ below $q^0$. This latest result in turn implies that the magnetization density is a conserved quantity. This is favorable to the application of the theory to the actinide $5f$ systems, though this result does not completely justify it. It is important to understand how the result reconciles with the two theoretical studies. Haslbeck {\it et al.} also reported that the critical exponents [$\beta$ = 0.32(1), $\gamma$ = 1.23(3), and $\nu$= 0.63(2)] determined by them are similar to those ($\beta$ = 0.325, $\gamma$ = 1.241, and $\nu$ = 0.630) in the 3D Ising model\cite{guillou}. The value of $\gamma$ (= 1.23) is different from that ($\gamma$ = 1.0) determined in our previous study\cite{tateiwa1}. Our result is consistent with that in the previous neutron scattering study\cite{huxley3}. The reason for this discrepancy is not clear.

\subsection{Possible reasons for absence of the superconductivity in URhSi}
We discuss the absence of the superconductivity in URhSi and future prospects for the study of uranium ferromagnets. At first, we introduce our study on UGe$_2$\cite{tateiwa4}. We measured the dc magnetization under high pressure and determined the pressure dependence of the characteristic energy of the longitudinal spin fluctuations $T_0$. We have found a clear correlation between $T_{\rm sc}$ and $T_0$ in UGe$_2$. Our result suggests that the superconductivity is mediated by the spin fluctuations developed at the phase boundary of FM1 and FM2 in UGe$_2$. The correlation between the two quantities has been discussed in high-$T_{\rm c}$ cuprate and heavy-fermion superconductors\cite{moriya7,moriya9,nksato9}. The importance of the longitudinal spin fluctuations for the $p$-wave superconductivity in the ferromagnetic state has been theoretically pointed out as mentioned in the Introduction\cite{fay,roussev,kirkpatrick,wang}. The values of $T_0$ in URhGe and URhSi are similar as shown in Table II. Furthermore, this study suggests the similarity in the critical behavior of the magnetization between URhSi, URhGe, and UGe$_2$\cite{tateiwa1}. Then, one could expect the superconductivity in URhSi. However, it has not been observed at low temperatures down to 40 mK in URhSi according to Ref. 22. We discuss the absence of the superconductivity in URhSi from two viewpoints. 

  
(i) Generally, it is difficult to grow the high-quality single crystal of UTX ferromagnets with the orthorhombic TiNiSi-type crystal structure. Here, $T$ is a transition $d$ metal. $X$ is a $p$-block element\cite{sechovsky}. The unconventional non $s$-wave superconductivity in strongly correlated electron systems is sensitive to a small amount of nonmagnetic impurities\cite{pfleiderer0,stewart2}. The high sensitivity of $T_{\rm sc}$ in URhGe to the electronic mean-free path was analyzed with the Abrikosov-Gorkov model assuming that the mean free path is proportional to the inverse of the residual resistivity ratio (1/RRR)\cite{huxley2,hardy2}. One possibility is that the quality of samples of URhSi studied so far are not enough for the appearance of the superconductivity.  It is necessary to grow the high-quality single crystal of URhSi in order to check the appearance of the superconductivity at low temperatures. 

 (ii) We discuss this issue from the recent uniaxial stress experiment on URhGe\cite{braithwaite}. The uniaxial stress $\sigma$ applied along the magnetic hard $b$ axis enhances $T_{\rm sc}$, with a merging of the low- and high-field superconducting states in URhGe\cite{braithwaite}. The value of $T_{\rm sc}$ in URhGe increases from 0.4 K at ambient pressure to 0.8 K at $\sigma$ $\sim$ 0.6 GPa applied along the $b$ axis. Meanwhile, the ferromagnetic transition temperature $T_{\rm C}$ shows only a 10 \% decrease. There is no report for the dc magnetization under the uniaxial pressure. We speculate that the value of $T_0$ does not change significantly. The relation between $T_{\rm sc}$ and $T_0$ under uniaxial stress in URhGe may be different from approximately linear relations in the high-$T_{\rm c}$ cuprate and heavy-fermion superconductors\cite{moriya7,moriya9,nksato9}. It is interesting to note that the relation in FM1 of UGe$_2$ is expressed as ${T_{\rm sc}}{\,}{\propto}{\,}({T_0})^{\alpha}$ with ${\alpha}$ = 2.3 $\pm$ 0.1\cite{tateiwa4}. $T_{\rm sc}$ is very sensitive to $T_0$.  We suggest one possibility: $T_0$ is not a sole parameter that determines $T_{\rm sc}$ in the uranium ferromagnetic superconductors.

The magnetic susceptibility ${\chi}_b$ in a magnetic field along the magnetic hard $b$ axis increases with increasing uniaxial stress applied along the same direction in URhGe\cite{braithwaite}. The increase in $T_{\rm sc}$ of the compound may be related to enhanced transverse magnetic fluctuations along the hard $b$ axis as theoretically shown\cite{mineev2}. Mineev derived a theoretical expression of $T_{\rm sc}$ in a BCS-type formula $T_{\rm sc}{\,}={\,}{\varepsilon}{\,}{\rm exp}(-1/g)$ for a two-band superconducting state in orthorhombic systems where the coupling between the two components of the equal spin-triplet $p$-wave order parameter is taken into account\cite{mineev2}. The coupling constant $g$ is expressed as follows:
    \begin{eqnarray}
g = {({g_{1x}^{\uparrow}}+{g_{1x}^{\downarrow}})\over 2}+ {\sqrt{{({g_{1x}^{\uparrow}}-{g_{1x}^{\downarrow}})^2\over 4}+{g_{2x}^{\uparrow}}{g_{2x}^{\downarrow}}}}
   \end{eqnarray}
 Here, ${g_{1x}^{\uparrow}}$ and ${g_{1x}^{\downarrow}}$ are coupling constants for intraband pairing, and ${g_{2x}^{\uparrow}}$ and ${g_{2x}^{\downarrow}}$ for interband pairing. ${g_{1x}^{\uparrow}}$ and ${g_{1x}^{\downarrow}}$ are proportional to the magnetic susceptibility along the magnetic easy $c$ axis. The intra-band pairing is driven by the longitudinal spin fluctuations whose energy scale can be estimated from $T_0$. ${g_{2x}^{\uparrow}}$ and ${g_{2x}^{\downarrow}}$ are determined by the difference of the magnetic susceptibilities in the magnetic hard $b$ and $a$ axes (${{\chi}}_b{\,}-{\,}{\chi}_a$). Thus, ${g_{2x}^{\uparrow}}{g_{2x}^{\downarrow}}$ is proportional to $({{\chi_b}}-{{\chi}_a})^2$. Note that ${g_{2x}^{\uparrow}}{g_{2x}^{\downarrow}}$ = 0 in tetragonal systems. The strength of the anisotropy in the magnetic susceptibilities along the magnetic hard axes increases $T_{\rm sc}$. ${\chi}_b$ is about 5 times larger than ${\chi}_a$ in the ferromagnetic state of URhGe at ambient pressure\cite{hardy}. The enhancement of $T_{\rm sc}$ in the uniaxial stress applied along the $b$ axis can be understood as the result of the increasing transverse fluctuations along the magnetic hard $b$ axis. On the other hand, the degree of the anisotropy in the magnetic susceptibilities ${\chi}_b$ and ${\chi}_a$ is at most two in the ferromagnetic state of URhSi as shown in Fig. 3(b). The contribution to raise $T_{\rm sc}$ by this mechanism may be smaller than that in URhSi.

 It is not clear which pairing mechanism (the intraband or interband one) takes a dominant role for the appearance of the superconductivity at ambient pressure in this system. It is difficult to make quantitative estimates of contributions from the two mechanisms to $T_{\rm sc}$. If the intraband pairing mediated by the longitudinal spin fluctuations is dominant, the superconductivity with the similar value of $T_{\rm sc}$ to that in URhGe could appear in the high-quality sample of URhSi since the values of $T_0$ in URhSi and URhGe are similar. Meanwhile, if the interband pairing mediated by the transverse fluctuations plays a certain role in the superconductivity, the $T_{\rm sc}$ value in URhSi could be smaller than that of URhGe even if the quality of samples is improved. This is because the strength of the transverse fluctuations in URhSi is smaller than that in URhGe as discussed above. It is necessary to grow the high-quality single-crystal sample of URhSi to settle this problem. 

It has long been thought that the $p$-wave ferromagnetic superconductivity driven by the longitudinal spin fluctuations appears in the uranium ferromagnetic superconductors where the ferromagnetism can be described with the 3D Ising model. The importance of the longitudinal spin fluctuations for the ferromagnetic superconductivity has been stressed in the theoretical studies\cite{fay,roussev,kirkpatrick,wang}. Our study for the critical behavior of the magnetization in UGe$_2$ and URhGe suggests that the 3D Ising universality class is not appropriate to describe the ferromagnetic phase transition in the uranium ferromagnetic superconductors\cite{tateiwa1}. The pairing mechanism other than that driven by the longitudinal spin fluctuations might take an important role for the uranium ferromagnetic superconductors. The improvement of the sample quality in URhSi would provide good opportunity to make further progress toward a complete understanding of the uranium ferromagnetic superconductors.


\section{SUMMARY}
The critical behavior of the magnetization in uranium ferromagnet URhSi has been studied around its ferromagnetic transition temperature $T_{\rm C}{\,}{\sim}{\,}$10 K. We have analyzed the magnetic data with a modified Arrott plot, a Kouvel-Fisher plot, the critical isotherm analysis, and the scaling analysis in order to determine the critical exponent $\beta$ for the temperature dependence of the spontaneous magnetization $M_{\rm s}$ below $T_{\rm C}$, $\gamma$ for the magnetic susceptibility $\chi$, and $\delta$ for the magnetization isotherm at $T_{\rm C}$.  We determine the values of the critical exponents as $\beta$ = 0.300 $\pm$ 0.002, ${\gamma}'$ = 1.00 $\pm$ 0.02 for $T$ $<$ $T_{\rm C}$, $\gamma$ = 1.03 $\pm$ 0.02 for $T_{\rm C}$ $<$ $T$, and $\delta$ = 4.38 $\pm$ 0.04 from the scaling analysis and the critical isotherm analysis. The Widom scaling law ${\delta}{\,}={\,}1+{\,}{\gamma}/{\beta}$ is fulfilled with the obtained critical exponents. The ferromagnetic state in URhSi has strong uniaxial magnetic anisotropy. However, the obtained critical exponents differ from those in the 3D Ising model ($\beta$ = 0.325, $\gamma$ = 1.241, and $\delta$ = 4.82) with short-range exchange interactions. The values of the exponents $\beta$ and $\gamma$ in URhSi are similar to those in uranium ferromagnetic superconductors UGe$_2$ and URhGe, and uranium ferromagnets UIr and U(Co$_{0.98}$Os$_{0.02}$)Al. We suggest that the universality class of the ferromagnetic transition in URhSi belong to the same one for the uranium ferromagnetic superconductors and the uranium ferromagnets. In these uranium ferromagnets, the magnetic susceptibility $\chi$ shows a mean-field-theory-like behavior (${\chi}^{-1}$ $\propto$ $T$) around $T_{\rm C}$ but the critical exponents $\beta$ are close to those of 3D ferromagnets. The anomalous critical exponents in the uranium ferromagnets cannot be explained with previous theoretical studies for critical phenomena. This conventional critical behavior of the magnetization may reflect peculiar features in the ferromagnetism of the $5f$ electrons where the superconductivity could appear. The absence of the superconductivity in URhSi is discussed from several viewpoints. We suggest that a further progress could be expected through the improvement of the sample quality in URhSi for a deeper understanding of the ferromagnetic superconductivity in the uranium ferromagnets.

 \section{ACKNOWLEDGMENTS}
This work was supported by Japan Society for the Promotion of Science (JSPS) KAKENHI Grant No. JP16K05463. 

\bibliography{apssamp}

\end{document}